\documentclass[aps,prl,twocolumn,groupedaddress,showpacs]{revtex4}
\usepackage{graphicx}
\usepackage{dcolumn}
\usepackage{amsmath}
\begin{document}

\title{Neutron-Proton Analyzing Power at 12 MeV \\and
Inconsistencies in Parametrizations of Nucleon-Nucleon Data}

\author{R.T. Braun}
\author{W. Tornow}
\author{C.R. Howell}
\author{D.E. Gonz{\'a}lez Trotter}
\author{C.D. Roper}
\author{F. Salinas}
\author{H.R. Setze}
\author{R.L. Walter}
\affiliation{\textup{Department of Physics, Duke University,
  Durham, North Carolina   27708-0308\\
  Triangle Universities Nuclear Laboratory, Durham,
  North Carolina 27708-0308}}
\author{G.J. Weisel}
\affiliation{\textup{Department of Physics, 
Penn State Altoona, Altoona, PA 16601}}

\begin{abstract}
We present the most accurate and complete data set for the analyzing 
power $A_y(\theta)$ in neutron-proton scattering. 
The experimental data were 
corrected for the effects of multiple scattering, both in 
the center detector and in the neutron detectors.
The final data at $E_n$ = 12.0 MeV deviate considerably from the 
predictions of nucleon-nucleon phase-shift analyses and potential models.  
The impact of the new data on the value of the charged pion-nucleon 
coupling constant is discussed in a model study.  
\end{abstract}


\maketitle

\section{INTRODUCTION}

There are reasons why low-energy ($E_N \le$ 20 MeV) Nucleon-Nucleon ($NN$) 
scattering data might appear to be of limited use in constraining $NN$ 
phase-shift analyses (PSAs) \cite{Sto93,Arn02} and potential 
models (PMs) \cite{Sto94,Wir95,Mac96,Ent03,Epe05}.
For one thing, the deuteron bound-state properties already provide a fairly 
stringent constraint for any $NN$ PM, and might seem sufficient.
For another, low-energy scattering data can provide constraints only for the 
lower partial wave $NN$ interactions and, even then, can not 
determine individual partial waves.
For example, the low-energy analyzing power, $A_y(\theta)$, is governed 
by the three angular momentum $L$ = 1 interactions, $^3P_0$, $^3P_1$, 
and $^3P_2$.  
Although $NN$ data provide constraints on the $^3P$ phase 
shifts taken together, it cannot determine each parameter 
unambiguously \cite{Tor99}.  

Despite the very small magnitude of $NN$ $A_y(\theta)$, its importance
derives from the fact that it is possible experimentally 
to measure such data to great precision. 
As a result, $NN$ $A_y(\theta)$ can provide a crucial test of our 
understanding of the $NN$ interaction and of the nuclear force in general.

Perhaps the most important controversy surrounding current $NN$ 
interaction models concerns the pion-nucleon coupling constant,  
$g^2_{\pi}/4\pi$. 
In Ref.~\cite{Mac01}, Machleidt and Slaus point out that low-energy
proton-proton ($p$-$p$) $A_y(\theta)$ data are very sensitive to the 
neutral pion-nucleon coupling constant, implying a value of 
$g^2_{\pi^0}/4\pi \leq 13.4$ (see also Ref.~\cite{Mac00}). 
At the same time, the correct description of the quadrupole moment
of the deuteron and low-energy neutron-proton ($n$-$p$) $A_y(\theta)$ data
requires meson-exchange based $NN$ potential models to have values for the 
neutral and charged pion-nucleon coupling constants
$g^2_{\pi^0}/4\pi$ and $g^2_{\pi^\pm}/4\pi$, respectively, of 14.0 or larger.  
The latter finding is clearly inconsistent
with the results of the Nijmegen group's NI93 PSA  
($g^2_{\pi^0}/4\pi = 13.47 \pm 0.11$ and 
$g^2_{\pi^{\pm}}/4\pi = 13.54 \pm 0.05$ \cite{Sto93b}) 
and of the VPI group 
($g^2_{\pi^0}/4\pi = 13.3$ and $g^2_{\pi^{\pm}}/4\pi = 13.9$ 
from $NN$ scattering \cite{Arn94b,Arn95} and 
$g^2_{\pi^0}/4\pi = 13.75 \pm 0.15$ from $\pi^{\pm}p$ scattering \cite{Arn94a}).

In principle, this inconsistency can be reduced by assuming a
charge-splitting of the pion-nucleon coupling constant, {\em i.e.}, one could 
assume that the neutral pion couples to the nucleon with a different
strength than the charged pion.  
In Refs.~\cite{Mac01,Mac00} it was shown that the combination of 
$g^2_{\pi^0}/4\pi = 13.6$ and $g^2_{\pi^{\pm}}/4\pi = 14.4$
creates a sufficiently large value for the quadrupole moment
of the deuteron, and reproduces the low-energy 
$p$-$p$ $^{3}P_0$ phase shifts.  

At the same time that the analyses of Refs.~\cite{Mac01,Mac00} were performed,
there were indications that $n$-$p$ differential cross-section 
data at intermediate energies favored a larger value for $g^2_{\pi^{\pm}}/4\pi$.
On this, see Ref.~\cite{Blo00} for a comprehensive overview
of recent determinations of $g^2_{\pi}/4\pi$ and especially Ref.~\cite{Rah98}, 
which quotes $g^2_{\pi^{\pm}}/4\pi = 14.50 \pm 0.26$  
obtained from $n$-$p$ differential cross section data at $E_n$ = 162 MeV.
However, the recent $n$-$p$ differential cross section data 
obtained at IUCF at 194 MeV \cite{Sar05} do not support this 
larger value of $g^2_{\pi^{\pm}}/4\pi$.

Although it seems likely that there is no significant charge splitting
in the pion-nucleon coupling constants at intermediate energies, 
the question remains unresolved at low energies. 
On the one hand, the theoretical models used to account for the charge 
dependence of the singlet $NN$ scattering lengths, $^{1}S_0$, do not allow 
for any large charge splitting of $g^2_{\pi}/4\pi$.
On the other hand, many low-energy data suggest a significant 
charge splitting.  
We report here on the results of a new $n$-$p$ $A_y(\theta)$ experiment 
carried out at $E_n$ = 12.0 MeV utilizing improved 
data-taking and data-analysis techniques.  
For references to previous $n$-$p$ $A_{y}(\theta)$ measurements 
see Refs.~\cite{Hol88,Tor88,Wei92}.
Our results confirm the inconsistencies between low-energy analyzing 
power and available theoretical models of the $NN$ interaction. 

\section{EXPERIMENTAL SETUP}

\begin{figure}[htb]
\includegraphics[height=2.6in]{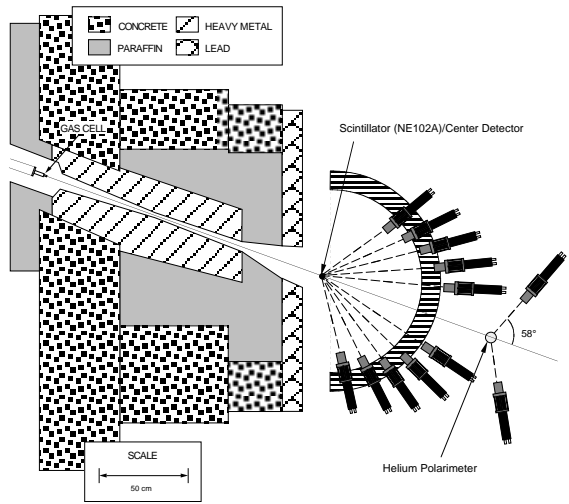}
\caption{Experimental setup for $n$-$p$ $A_y(\theta)$ measurements in 
TUNL's Shielded Neutron Source area.}
\label{fig:setup}
\end{figure}

The experimental setup is shown in Fig.~\ref{fig:setup}.  Polarized neutrons
with mean energy of 12.0 MeV and total energy spread of about 400 keV were 
produced via the polarization-transfer reaction 
$^2$H($\vec{d}$,$\vec{n}$)$^3$He at $0^{\circ}$.  The polarized deuteron beam 
was accelerated to $E_d$ = 9.40 MeV and entered through a 4.6 $\mu$m Havar foil
into a small (3.14 cm long, 0.48 cm radius) gas cell 
filled with 7.8 atm of deuterium gas and capped with a 0.1 cm thick gold beamstop.  
The gas cell was mounted inside a 1.8 m-thick wall made of concrete,
paraffin, iron, copper, and lead, to shield the neutron detectors from the 
direct flux of the neutron source.  Typical deuteron
beam intensities on target were 1.5 $\mu$A and typical values for the deuteron
vector polarization $|p_z|$ were 0.65.  The polarized neutrons
produced at $0^{\circ}$ relative to the incident deuteron beam passed 
through a collimation system to produce a rectangular shaped neutron beam 
at the position of the proton-containing active 
target labelled ``Center Detector'' in Fig.~\ref{fig:setup}.
The center detector (CD) consisted of an upright cylinder made of the plastic
scintillator material NE102A with dimensions 1.9 cm diameter and 3.8 cm height.
The CD was located at a distance of 172 cm from the neutron source
and was mounted via a short light guide onto a 5 cm diameter photomultiplier 
tube (PMT).  

Neutrons scattered to the left or right were detected by five pairs of 
neutron detectors positioned symmetrically relative to the
incident neutron beam direction in the horizontal scattering plane.  
The neutron detectors (NDs) were filled with the liquid scintillator 
material NE213.  These detectors had excellent neutron-gamma pulse-shape
discrimination capabilities and had an active volume of 
4.3 cm wide, 11.9 cm high and 7.5 cm deep.  They were viewed by 5 cm diameter
PMTs through 0.5 cm thick Pyrex glass windows and 7.5 cm long light guides.
The neutron detectors were mounted onto (low-mass) 30 cm high stands and placed
on an aluminum ring surrounding the CD.  
The center-to-center distance between the CD and the
neutron detectors ranged from 45 cm to 70 cm depending on scattering angle.
The angular separation between the neutron detectors was $12^{\circ}$ (lab).  
In order to cover the angular range from $\theta_{lab} = 16^{\circ}$ to 
72$^{\circ}$ in 4$^{\circ}$ steps, three settings of the five detector
pairs were required.
The absolute magnitude of the neutron polarization was measured with a neutron
polarimeter located downstream of the $n$-$p$ scattering arrangement.  The 
polarimeter consisted of a $^4$He gas scintillator pressurized
to 100 atm (95\% He, 5\% Xe) and a pair of neutron detectors positioned at
$\theta_{lab} = 58^{\circ}$, which 
were identical to those used for $n$-$p$ scattering.
In order to reduce instrumental asymmetries for the $n$-$p$ and $n$-$^4$He
measurements, the deuteron vector polarization
$p_z$, and therefore the neutron polarization, was flipped at a frequency of 
10 Hz (between up and down relative to the horizontal scattering plane).
The $n$-$p$ and $n$-$^4$He data were accumulated simultaneously in six 
runs, each lasting about 250 data-taking hours.  

The data-acquisition electronics recorded the center detector pulse height 
(CDPH), the neutron time of flight (NTOF) between the CD and the NDs, and 
spectra for each ND, including pulse-shape information. 
Since the energy of the scattered neutrons varied from $E_{n'}$ = 11.1 MeV at 
$\theta_{lab} = 16^{\circ}$ to $E_{n'}$ = 1.1 MeV at 
$\theta_{lab} = 72^{\circ}$, 
different hardware thresholds were used for the NDs.
In addition, three different gains
were used for the CD signals (using different dynodes).

\begin{figure}[htb]
\includegraphics[width=2.6in]{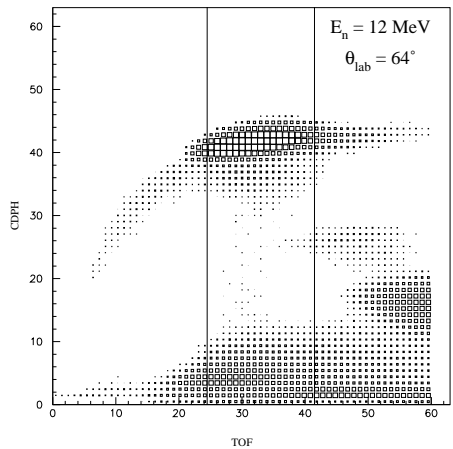}
\caption{2D spectrum of compressed CDPH versus NTOF for 
scattering to $\theta_{lab}$ = 64$^{\circ}$.  A tight NTOF gate was set around
the elastic neutron peak in order to remove as many background events 
as possible.}
\label{fig:2d}
\end{figure}

Software cuts were set on the CDPH and the pulse height in the NDs 
to eliminate pulses at the extreme ends of the spectra.
Gates were also set on the neutrons in the pulse-shape 
discrimination spectra and wide gates were set on the elastic peak
of the NTOF spectrum. 
All four of these cuts (identical for spin-up and spin-down spectra)
were used to generate two-dimensional (2D) 
spectra of CDPH versus NTOF, for scattering to the left and right NDs 
and for neutron spin up and spin down.
An example of such a spectrum is shown in Fig.~\ref{fig:2d} where the CDPH
scale has been temporarily compressed in order to fit within 64 channels.
Tight NTOF gates were set in these 2D spectra eliminating the tails of the peak,
as shown in Fig.~\ref{fig:2d},
in order to identify the elastic scattering events of interest
(again, identical for spin-up and spin-down spectra).
These new NTOF gates were used to sort the final CDPH spectra 
(now in their full 512-channel resolution) 
corresponding to each neutron detector and spin state.  
The CDPH spectra were used 
to determine the $n$-$p$ yields and scattering asymmetries, after 
applying the corrections described in the following section.
The above process was also followed to sample the accidental
({\em i.e.,} time uncorrelated) background by using an NTOF cut 
located at times shorter than the gamma peak.  
The accidental background proved to be extremely small.  

\section{DATA ANALYSIS}

After the sorting procedure described above 
and the subtraction of the accidental events, the data still
contained a number of finite-geometry and multiple-scattering effects.  
To remove these effects, Monte-Carlo calculations were performed to 
simulate the experiment.  Two effects are due exclusively to the finite
size of the center detector (CD) and the neutron detectors (NDs) and 
have a slight effect on single-scattering events. 
First, because there is a range of angles subtended by each detector 
set at each nominal angle and because the cross section
of $n$-$p$ elastic scattering varies over this range, we must report an 
effective angle.  These were calculated by our code and are listed in the 
first column of Table \ref{tab:ay}.  This effect is small; the largest 
shift is no more than a half of a degree.  
The second finite-geometry effect concerns the value of $A_y{(\theta)}$ 
itself, again due to the range of angles subtended by each ND.  
Effective $A_y{(\theta)}$ values were calculated by our Monte-Carlo code 
and these were compared to the values from the code's library.  The 
ratio between these two values was then applied to the data.  
Once again, the correction is small; only the first four angles had 
corrections that were larger than the uncertainty of the calculation 
(about 0.00012).
 
In addition to elastic scattering, multiple scattering events occur in the CD. 
About 50\% (depending on ND angle) of these events were eliminated as a 
result of the neutron time of flight (NTOF) gate. 
Nevertheless, the center detector pulse height (CDPH) spectrum contained 
multiple scattering events amounting to approximately 2\% of all single 
scattering events.
Our Monte-Carlo simulation showed that the only 
significant processes were those due to double scattering, specifically  
neutron double scattering from hydrogen ($^1$H-$^1$H), neutron scattering 
from hydrogen and subsequent scattering from carbon ($^1$H-$^{12}$C), and 
neutron scattering from carbon and subsequent scattering from hydrogen 
($^{12}$C-$^1$H).  In performing these calculations, we used 
complete libraries of cross-section and polarization data for both 
$n$-$^1$H and $n$-$^{12}$C scattering.  We will return to the subject 
of the $n$-$^{12}$C library in our discussion of the PDE correction.

We also removed edge-effect events from the data, which result
when recoil protons leave the CD before depositing their full energy.  
Along with the double scattering events, these counts elongate the tails of
the CDPH peak, especially to the left (low-energy) side.  

\begin{figure}[htb]
\includegraphics[width=2.8in]{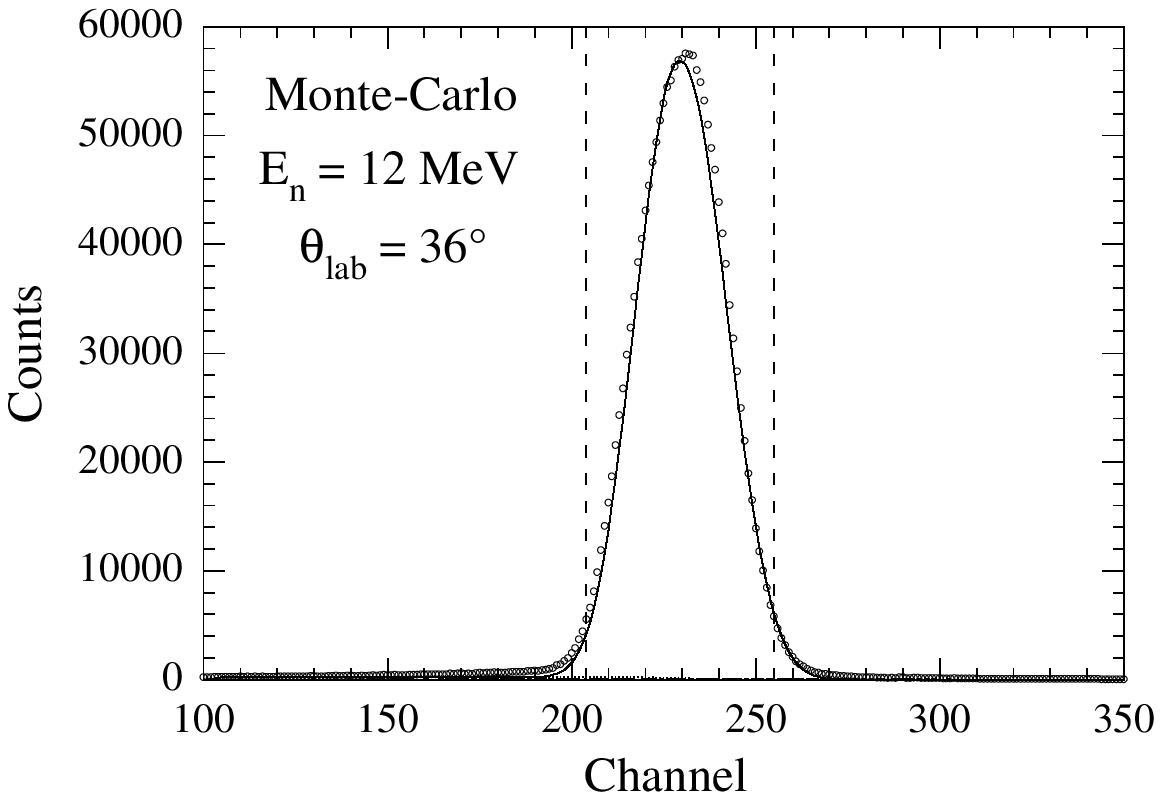}
\includegraphics[width=2.8in]{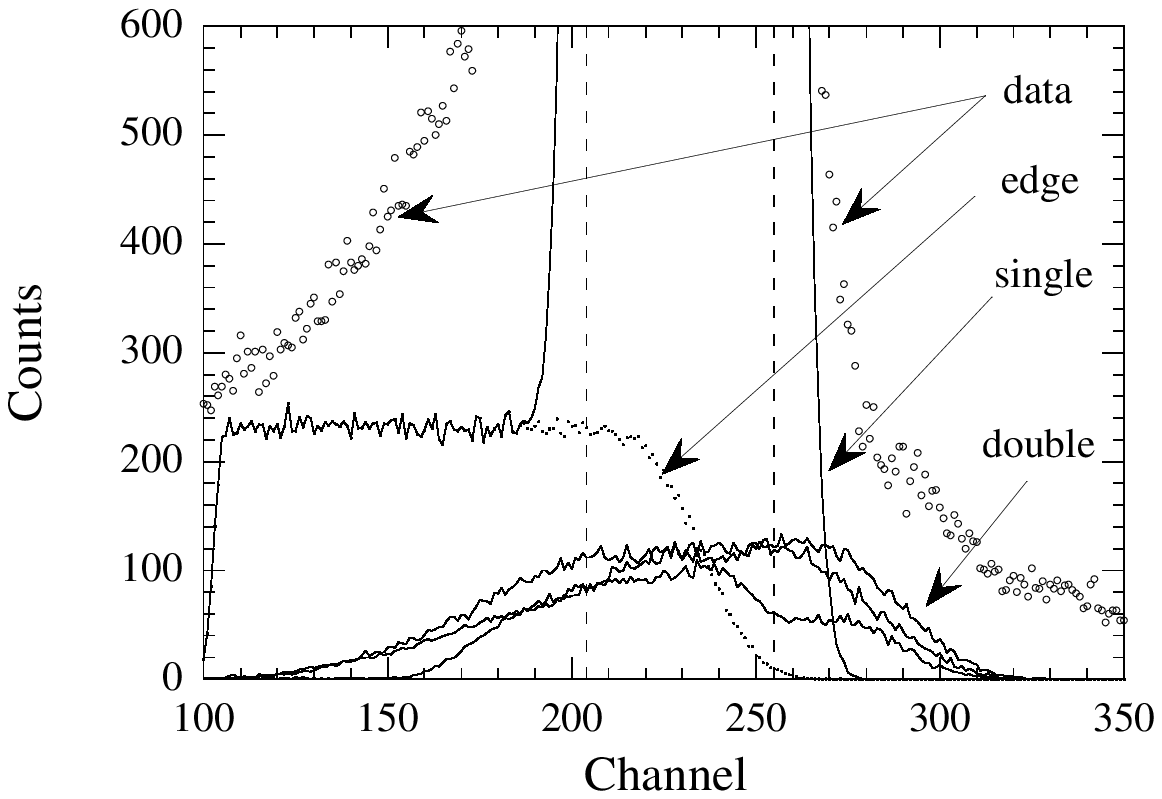}
\includegraphics[width=2.8in]{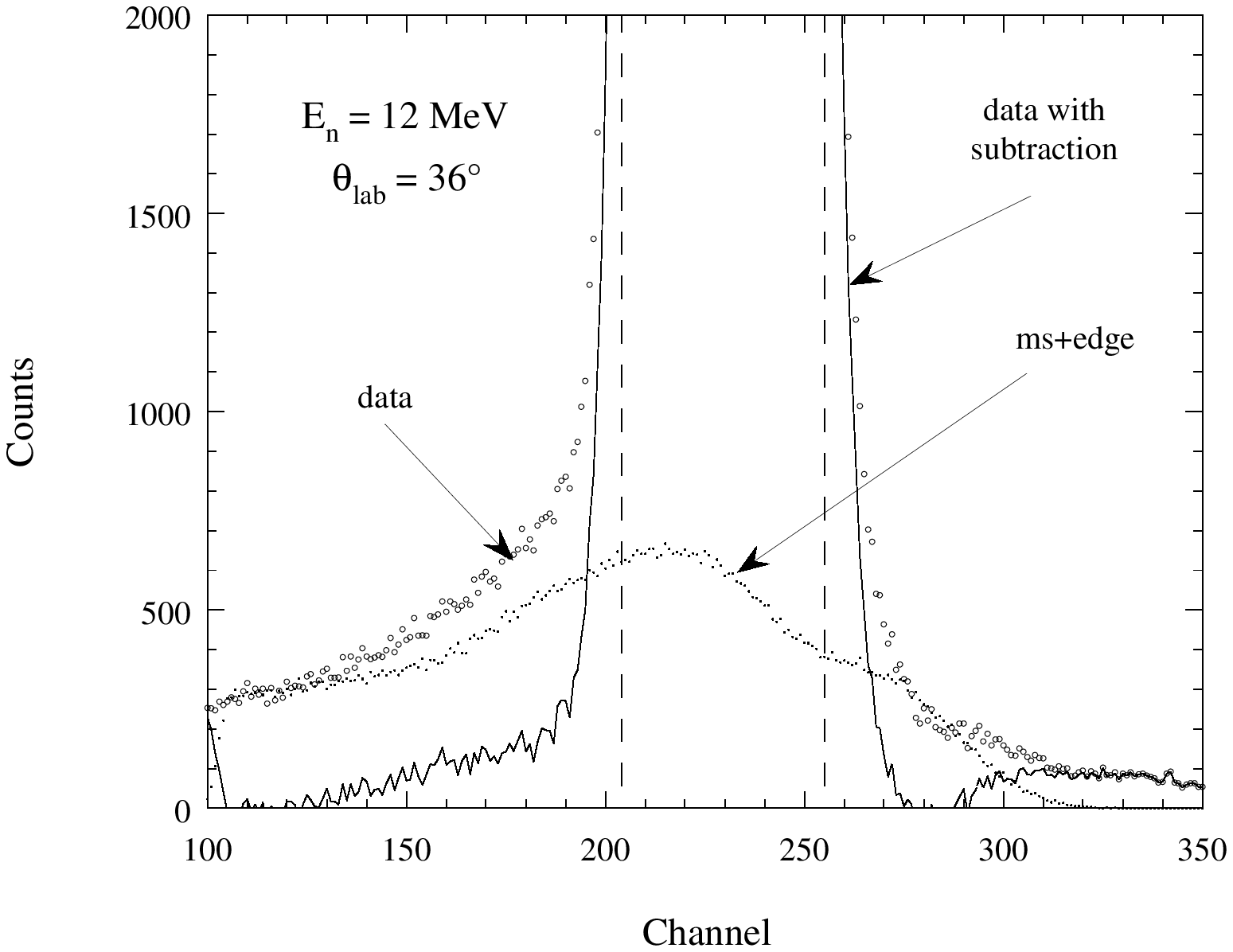}
\caption{The top panel shows a comparison of calculated (solid curve) 
and measured (dots) CDPH spectrum for scattering 
to $\theta_{lab}$ = 36$^{\circ}$.  
The middle panel shows an expanded view with focus on calculated 
multiple-scattering and edge effect contributions.  
The bottom panel shows an additional expanded view focusing
on the remaining background.  See text for details.}
\label{fig:prph}
\end{figure}

A sample CDPH spectrum is shown in Fig.~\ref{fig:prph} (top panel).
The solid curve represents our Monte-Carlo simulation for a scattering
angle of $\theta_{lab} = 36^{\circ}$, while the small open circles show the 
experimental data.  A greatly expanded view is shown in the middle panel,
where the open circles again indicate the experimental data.  
The curves labeled ``double'' are the calculated double scattering 
contributions $^{1}$H-$^{1}$H, $^{1}$H-$^{12}$C, and $^{12}$C-$^{1}$H.  
The dotted curve labled ``edge'' is the calculated pulse-height 
distribution due to edge effects. 
Finally, the curve labeled ``single'' is the calculated single 
scattering contribution (plus the edge effects).  
All of the calculations are normalized to the data.  

Another expanded view is displayed in the bottom panel of Fig.~\ref{fig:prph}. 
Here, ``data with subtraction'' shows the data after the removal of all
counts due to double scattering and edge effects (labeled ``ms+edge'').  
Even after this subtraction, a small background remains, 
amounting to about 0.3\% of the single-scattering events.
A number of fits were used to estimate this remaining background, 
ranging from a linear fit between channel numbers 150 and 350 to a 
parabolic fit between channel numbers 180 and 280.  Due to the 
smallness of the remaining background, the asymmetry proved to 
be independent of our background choice, within statistical uncertainties.
We also concluded that the background was unpolarized.  
For all ND angle settings we approximated the remaining 
background by a linear function connecting the left and right sides
of the CDPH peak (for example, in Fig.~\ref{fig:prph} from channel
180 to 280).  The remaining background seen above channel 290 in the 
bottom panel of Fig.~\ref{fig:prph} is due to cross-talk effects
between two adjacent detectors, specifically neutron scattering from
the detector positioned at a larger scattering angle 
(and shorter distance from the CD) to the detector of interest.

Three sets of gates were used to calculate the yields and asymmetries, 
at 10\% (shown in Fig.~\ref{fig:prph} by the dashed lines), 
30\%, and 50\% of the CDPH peak maximum.  
In Fig.~\ref{fig:prph}, this is done for $N_L^{\uparrow}$, 
for spin up scattering to the left ND at $\theta_{lab} = 36^{\circ}$.  
Similarly, the yields
$N_R^{\uparrow}$, $N_L^{\downarrow}$, and $N_R^{\downarrow}$ were obtained to
calculate the asymmetry $\epsilon = (\alpha - 1)/(\alpha + 1)$ with 
$\alpha = \sqrt{\frac{N_L^{\uparrow}}{N_R^{\uparrow}}
\frac{N_R^{\downarrow}}{N_L^{\downarrow}}}$. 
The nominal gates were the 30\% set.
The other two gates (the 10\% and 50\% set)
were used to check on the appropriateness of the background
subtraction.  Within statistical uncertainty, the results for $\epsilon$ 
proved to be independent of the choice of the gate width.

In order to extract the $n$-$p$ $A_y(\theta)$ from the measured asymmetry
$\epsilon (\theta)$, the neutron polarization $p_y^n$ must be known.
For this purpose the $n$-$^4$He asymmetry data acquired with the neutron
polarimeter referred to above were processed and analyzed in the same way
as the $n$-$p$ asymmetry data.  In this case the $^4$He recoil pulse height
in the high-pressure gas scintillator plays the role of the CDPH in the
plastic scintillator used for the $n$-$p$ asymmetry measurements. The neutron
polarization was obtained from $\epsilon_{He}(58^{\circ}) = (\alpha_{He} - 1)/
(\alpha_{He} + 1) = \bar{A_y}(58^{\circ})p_y^n$, where $\alpha_{He}$ is
defined as above.  Here, the effective
analyzing power $\bar{A_y}(58^{\circ})$ for $n$-$^4$He scattering at
$E_n$ = 12.0 MeV was calculated for the present neutron polarimeter geometry via
Monte-Carlo calculations.  The $n$-$^4$He phase shifts of Stammbach
and Walter \cite{Sta72} were used.  All of the relevant multiple 
scattering processes 
were included.  We obtained $\bar{A_y}(58^{\circ}) = -0.554 \pm 0.008$, where 
the uncertainty is mainly of a systematic nature reflecting the uncertainty 
associated with the $n$-$^4$He phase shifts.  The average neutron polarization
was $p_y^n = 0.563 \pm 0.008$.

At such a high level of precision, a subtle systematic
effect comes into play, which does not cancel by reversal of the neutron
polarization.  This is the polarization 
dependent efficiency (PDE) \cite{Hol88} of the neutron detectors. 
The NDs contain hydrogen and carbon in the ratio of 1.21:1.
The double scattering process $^{12}$C-$^1$H in the NDs, which accounts for
about 10\% of the total neutron detection efficiency,
is sensitive to the $n$-$^{12}$C $A_y(\theta)$.  If the 
$n$-$^{12}$C $A_y(\theta)$ is not constant over the range of neutron energies
$E_{n'}$ seen by a particular ND, an instrumental asymmetry will occur.
Typical values for $\Delta E_{n'}$ are 800 keV.  A realistic correction for 
this effect requires a detailed knowledge of the $n$-$^{12}$C $A_y(\theta)$,
especially in the resonance region of the $n$-$^{12}$C total cross section
between 2.0 and 8.5 MeV neutron energy.  In this energy regime the 
$n$-$^{12}$C $A_y(\theta)$ changes rapidly and therefore causes sizeable
PDE effects.  

All of the post-1985 $n$-$p$ $A_y(\theta)$ measurements
have been corrected for the PDE.  However, due to the lack
of a detailed $n$-$^{12}$C $A_y(\theta)$ database, especially at
low energies, the accuracy of the associated corrections was limited.  
In assembling our data library, we used the  
thirty-three $n$-$^{12}$C $A_y(\theta)$
angular distributions measured by Roper {\em et al.}~\cite{Rop05}
in the energy range from 2.2 to 8.5 MeV.  
From $E_n =$ 0 to 6.5 MeV, we used an R-matrix analysis  
by Hale \cite{Hal00}, which included the data from Ref.~\cite{Rop05}.
In certain regions (especially for forward angles and for neutron
energies between 3.5 and 4.5 MeV), the analysis of Ref.~\cite{Hal00} 
missed the $A_y(\theta)$ data slightly and we therefore substituted 
Legendre polynomial fits to the data of Ref.~\cite{Rop05} in these regions.
Between 6.5 MeV and 8.5 MeV, we used fits to the data of Ref.~\cite{Rop05} as 
well as the recent phase-shift analysis (PSA) of Chen and Tornow \cite{Che05}.
Above $E_n =$ 8.5 MeV, we used the Chen-Tornow PSA exclusively.
The new data by Roper {\em et al.}~and the analyses of Hale, Chen and Tornow 
improved the $n$-$^{12}$C $A_y(\theta)$
database considerably, making corrections for the PDE more reliable.

We ran our Monte-Carlo code for 20 separate legs, each leg of 
three million events, and each leg starting from a different random number.  
The PDE correction to the $A_y(\theta)$ data
was taken as the difference between the $A_y(\theta)$ result with 
polarization effects turned on in the NDs and the result with
the polarization turned off. 
The second column of Table \ref{tab:ay} lists our final PDE corrections.
Note that they vary greatly from one data point to the other due to the
pronounced resonance features in $n$-$^{12}$C scattering at low energy.  
It is important also to note that our present results agree well with 
the overall trend of the PDE corrections of Ref.~\cite{Wei92}, which
used a different Monte-Carlo code and a different database.  
Our reason for having much greater confidence in the present PDE results
is due to our extensive and detailed work in revising the data libraries, 
as outlined above.

The third column in Table \ref{tab:ay} summarizes our final results for 
$A_y(\theta)$ in $n$-$p$ scattering at $E_n$ = 12.0 MeV.  
Note the small overall uncertainty.  
The final results include uncertainties in $A_y(\theta)$ 
due to statistics, the measurement of beam polarization,
the multiple-scattering calculation, the PDE calculation, and the
remaining background (typically zero) all added in quadrature.  
The final uncertainties are about half of those of the previous 
TUNL $n$-$p$ $A_y(\theta)$ measurement at $E_n$ = 12.0 MeV \cite{Wei92}.  
This is partly due to the fact that the Atomic Beam Polarized Ion Source
used in the present study produced about four times the deuteron current
as the Lamb-Shift source used in the previous study.

\section{DISCUSSION}

\begin{figure}[htb]
\includegraphics[width=2.6in]{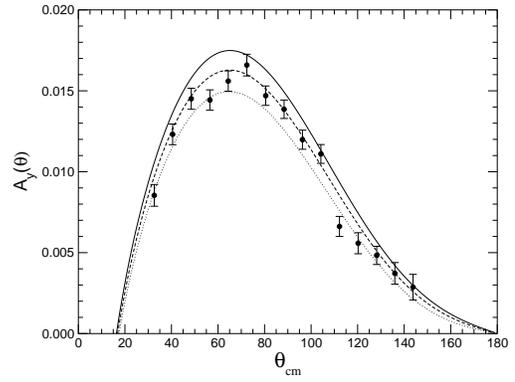}
\caption{Neutron-proton $A_y(\theta)$ data at $E_n = 12.0$ MeV in comparison 
to theoretical predictions.  The error bars associated with the 
data represent the overall uncertainty of the data with statistical and 
systematic uncertainties added in quadrature.
The solid curve is the Nijmegen NI93 PSA prediction.
The other curves are for the CD-Bonn based model study which varies the 
charged pion coupling constant.
Here, for $g^2_{\pi^0}/4\pi$, all three curves use 13.6.
For $g^2_{\pi^\pm}/4\pi$, the calculation using 13.6 coincides on this scale 
with the Nijmegen NI93 PSA result (solid curve); the dashed curve uses 
14.0 and the dotted curve 14.4.} 
\label{fig:data}
\end{figure}

Figure \ref{fig:data} shows the present $n$-$p$ $A_y(\theta)$ in comparison to
the $NN$ phase-shift analysis prediction (solid curve) of the Nijmegen 
group, NI93.  
Clearly, NI93 provides a larger $A_y(\theta)$ throughout the entire
angular distribution.  
The accuracy of the neutron polarization determined in the present work 
does not allow for a renormalization of the 
$A_y(\theta)$ data beyond the error bars given in Fig. \ref{fig:data}.  
Furthermore, the present 
$n$-$p$ $A_y(\theta)$ data are in good agreement with the trend
established by previous TUNL data where a different method was used for
determining $p_y^n$ \cite{Wei92}.

As we have pointed out in the introduction, the underlying $NN$ dynamics that 
characterize $A_y(\theta)$ precludes us from
extracting unambiguous information about the $^3P_j$ $NN$ interactions.  
However, we can conclude that the NI93 $NN$ PSA overestimates the $n$-$p$
$A_y(\theta)$ at $E_n$ = 12.0 MeV.  This statement is
of considerable importance considering the fact that most $NN$ potential 
model builders use the NI93 PSA results or the associated database
for determining
the free parameters of their models.  One has to conclude that
all the recent so-called high-precision $NN$ potential models overestimate
the $n$-$p$ $A_y(\theta)$ at low energies.  This observation has far-reaching
consequences for nuclear scattering systems with $A > 2$, which are much
more sensitive to the $^3P_j$ $NN$ interactions than the $NN$ 
system \cite{Tor98}.

Valuable information can be obtained from the present data if they are 
compared to variations of the theoretical predictions.  
Here we focus on the charged pion coupling constant \cite{Mac01,Mac00}.  
Figure \ref{fig:data} shows our data in comparison to three theoretical 
predictions based on the CD-Bonn $NN$ potential, which use three different 
values of the charged pion-nucleon coupling constant, $g^2_{\pi^\pm}/4\pi$. 
In these three models, only the S-wave NN interactions of CD-Bonn 
were refitted.  
All three predictions use the same neutral pion coupling constant, 
$g^2_{\pi^0}/4\pi = 13.6$.
The curve using $g^2_{\pi^\pm}/4\pi = 13.6$ is indistinguishable on this 
scale from the prediction of NI93 (solid curve).
The dashed curve in Fig.~\ref{fig:data} uses $g^2_{\pi^\pm}/4\pi = 14.0$ and 
the dotted curve uses $g^2_{\pi^\pm}/4\pi = 14.4$. 
The values of $\chi^2$ per degree of freedom associated with the solid, 
dashed and dotted curves are 6.0, 1.7, and 2.5, respectively.
Therefore, this model study confirms and 
puts on more solid ground the findings of Refs.~\cite{Mac01,Mac00} 
regarding low-energy $n$-$p$ $A_y(\theta)$
data and their demand for a charge splitting of $g^2_{\pi}/4\pi$. 

In summary, the present data represent the most accurate and complete
$n$-$p$ $A_y(\theta)$ angular distribution ever reported.
Our model study based on the CD-Bonn $NN$ potential model 
supports a substantial charge dependence of the pion-nucleon coupling constant.
Our results are inconsistent with the existing global $NN$ PSAs of the 
Nijmegen \cite{Sto93b} and VPI \cite{Arn94b,Arn95} groups 
and with high-precision $NN$ potential models.  
However, our results agree with inconsistencies previously noticed
between data and predictions for the $^3S_1$-$^3D_1$ mixing parameter
$\epsilon_1$ in $n$-$p$ scattering at low energies \cite{Tor02}
and also with requirements placed on the charged coupling constant by 
the quadrupole moment of the deuteron \cite{Mac01}.
Of course, it is possible that neither of these scenarios is the 
``correct'' one.  Perhaps the impasse comes because we are
at the point where the precision of our data and the development
of our ``low energy'' theoretical models has pushed the paradigm of 
meson-exchange based $NN$ potential models beyond its limits.

\begin{acknowledgments}
This work was supported in part by the U.S. Department of Energy, Office of
Nuclear Physics, under Grant No.~DE-FG02-97ER41033.  The authors would like
to thank R.~Machleidt for valuable contributions to this work.
\end{acknowledgments}


\begin{table}
\caption{Results of $n$-$p$ $A_y(\theta)$ experiment at $E_n$ = 12.0 MeV.}
\begin{tabular}{c|c|c}
$\theta_{c.m.}$ & PDE correction & final results\\
\hline
32.6 &0.00014 $\pm$ 0.00016 &0.00854 $\pm$ 0.00067 \\
40.5 &0.00004 $\pm$ 0.00016 &0.01231 $\pm$ 0.00064 \\
48.5 &0.00006 $\pm$ 0.00015 &0.01451 $\pm$ 0.00065 \\
56.5 &0.00013 $\pm$ 0.00013 &0.01443 $\pm$ 0.00063 \\
64.4 &-0.00005 $\pm$ 0.00015 &0.01560 $\pm$ 0.00063 \\
72.4 &0.00294 $\pm$ 0.00022 &0.01659 $\pm$ 0.00067 \\
80.5 &-0.00185 $\pm$ 0.00014 &0.01470 $\pm$ 0.00060 \\
88.4 &0.00019 $\pm$ 0.00017 &0.01386 $\pm$ 0.00057 \\
96.3 &0.00072 $\pm$ 0.00018 &0.01198 $\pm$ 0.00059 \\
104.2 &-0.00136 $\pm$ 0.00018 &0.01110 $\pm$ 0.00058 \\
112.2 &-0.00114 $\pm$ 0.00028 &0.00662 $\pm$ 0.00062 \\
120.2 &0.00108 $\pm$ 0.00029 &0.00558 $\pm$ 0.00065 \\
128.2 &0.00103 $\pm$ 0.00021 &0.00483 $\pm$ 0.00056 \\
136.0 &-0.00018 $\pm$ 0.00036 &0.00372 $\pm$ 0.00067 \\
143.8 &-0.00029 $\pm$ 0.00040 &0.00287 $\pm$ 0.00079 \\
\end{tabular}
\label{tab:ay}
\end{table}

\clearpage

\clearpage

\clearpage

\end{document}